\documentstyle[aps,prb,twocolumn,epsf,rotate]{revtex}

\begin{document}
%% The following two lines should be there when using 'twocolumn'.
\twocolumn[\hsize\textwidth\columnwidth\hsize\csname
@twocolumnfalse\endcsname

%%%%%%%%%%%%%%%%%%%%%%%%%%%%%%%%%%%%
%
\newcommand{\beq}{\begin{equation}}
\newcommand{\eeq}{\end{equation}}
\newcommand{\beqa}{\begin{eqnarray}}
\newcommand{\eeqa}{\end{eqnarray}}
%%%%%%%%%%%%%%%%%%%%%%%%%%%%%%%%%%%%%%%%%%%%%%%%%%%%%%%%%
\newcommand{\eqbreak}{
\end{multicols}
\widetext
\noindent
\rule{.48\linewidth}{.1mm}\rule{.1mm}{.1cm}
}
\newcommand{\eqresume}{
\noindent
\rule{.52\linewidth}{.0mm}\rule[-.1cm]{.1mm}{.1cm}\rule{.48\linewidth}{.1mm}
\begin{multicols}{2}
\narrowtext
}
%%%%%%%%%%%%%%%%%%%%%%%%%%%%%%%%%%%%%%%%%%%%%%%%%%%%%%%%%
%

\title{Extensive eigenvalues in spin-spin correlations:            
a tool for counting pure states in Ising spin glasses}

\draft

\author{Jairo Sinova$^{1,2}$, Geoff Canright$^{1,2}$, 
Horacio E. Castillo$^{3,4}$, and Allan H. MacDonald$^{2}$}
\address{$^{1}$Department of Physics,
University of Tennessee, Knoxville, Tennessee}
\address{$^{2}$Department of Physics,
Indiana University, Bloomington, Indiana}
\address{$^{3}$ CNRS-Laboratoire de Physique Th{\'e}orique de 
l'Ecole Normale Sup{\'e}rieure, Paris, France.}
\address{$^{4}$Department of Physics,
Boston University, Boston, Massachusetts}
\date{\today}
\maketitle

\begin{abstract}
  We study the nature of the broken ergodicity in the low temperature phase
  of Ising spin glass systems, using as a diagnostic tool 
  the spectrum of eigenvalues of the spin-spin correlation function.
We show that multiple
  extensive eigenvalues of the correlation matrix $C_{ij}\equiv\langle S_i  
  S_j\rangle$ occur if and only if there is replica symmetry breaking. 
We support our
  arguments with Exchange Monte-Carlo results for the infinite-range problem.
  Here we find multiple extensive eigenvalues in the RSB phase for $N \agt
  200$, but only a single extensive eigenvalue for phases with long-range
  order but no RSB. Numerical results for the short range model in four
  spatial dimensions, for $N\le 1296$, are consistent with the presence of a
  single extensive eigenvalue, with the subdominant eigenvalue behaving
  in agreement with expectations derived from the droplet model. 
  Because of  the small system sizes we cannot exclude the 
  possibility of replica symmetry breaking with finite size corrections in 
  this regime.   
\end{abstract}


%% The following line should be there when using 'twocolumn'.
\vskip2pc]

\section{Introduction}
\label{SEC:Intro}
In three spatial dimensions or higher
it is now accepted that Ising spin systems with random exchange interactions,
exhibiting disorder and frustration, undergo a transition from a paramagnetic
phase to a glass phase at a finite temperature.~\cite{B&Y}   %A103
By contrast, the nature of the glass phase at finite
dimensions is still a subject of much debate in the literature, with
two main competing points of view.  One follows       %A103
Parisi's solution of the Sherrington-Kirkpatrick (SK) model~\cite{SK} using a
replica symmetry breaking (RSB) ansatz.~\cite{Parisi} The Parisi solution
involves broken ergodicity of a subtler form than that found in a
conventional ferromagnet: configuration space is broken into many ergodic
regions, separated by energy barriers which diverge in the thermodynamic
limit.  Most of these regions---which we will also call pure states---are
unrelated to one another by any symmetry of the Hamiltonian.  However, in the
case of a Hamiltonian with global spin inversion symmetry, each of these
regions has an associated region related to it by global spin inversion, the
pair forming together what we will call a pure state pair (PSP).  This RSB
picture is almost certainly correct in the limit of infinite spatial
dimension; and it has been argued that the RSB picture also applies, more or
less unchanged, to the frozen phase of finite-dimensional spin glasses. The
other point of view is the `droplet' picture,~\cite{droplet} which, in sharp
contrast to the picture just described, postulates the presence of a single
PSP in the low-temperature phase.  In this paper we focus on the fundamental
difference between these two pictures, namely, the nature of the ergodicity
breaking in spin glasses---or, put more simply, the number of `valleys'
(ergodic regions) in the low-temperature phase.

The glass transition is characterized by the
Edwards-Anderson spin-glass order parameter
\begin{equation}
q=\frac{1}{N}[\sum_i \langle S_i\rangle^2]_{\rm av},
\label{EQ:qEA_def}
\end{equation}
which becomes nonzero [i.e., of $O(1)$] for $T<T_c$.  Here $S_i= \pm 
1$ is the spin at site $i$, $N$ is the number of spins in the system,
$[\ ]_{\rm av}$ indicates an average over disorder realizations and
$\langle \phantom{X} \rangle$ indicates a thermal average.  The order parameter %%G1016 
$q$ may also be obtained as the first moment of the overlap
distribution function
\begin{equation}
P(q)= \left[ \sum_{\alpha\beta} w^\alpha w^{\beta}
\delta(q-q^{\alpha\beta}) \right]_{\rm av}\,,
\label{EQ:Pq_def}
\end{equation}
where $\alpha$,$\beta$ label pure states, $q^{\alpha\beta} \equiv
\sum_i \langle S_i\rangle^\alpha \langle S_i\rangle^{\beta}/N$
indicates the overlap between the local magnetizations in two pure
states, and $w^\alpha$ is the thermodynamic weight of pure state
$\alpha$. In the thermodynamic limit, $P(q)$ is predicted to have very
different behaviors in the two pictures mentioned above. For a single
PSP, $P(q)$ approaches a pair of delta functions at $\pm q_{max}$, where
$q_{max}$ is the self-overlap of each pure state in the pair, and
$-q_{max}$ is the overlap between one pure state and the other. For the
diverging (countable) number of PSPs in the SK problem $P(q)$ is
nonzero over a finite interval $-q_{max} \le q \le q_{max}$ (this
property depends essentially on the fact that $P(q)$ is a disorder
averaged quantity). So far, the main tool for detecting
non-trivial broken ergodicity in finite dimension at nonzero
temperatures has been the scaling of the overlap distribution function
$P(q)$ at $q=0$ as a function of system size: for a single PSP it must
scale to zero, while it remains nonzero in the thermodynamic limit if
there is RSB.  Several numerical studies have suggested a behavior in
finite dimensions similar to the one at the mean field
level.~\cite{Young83,Marinari} Other studies using the Migdal-Kadanoff
approximation,~\cite{moore} and still others investigating the ground
state susceptibility to boundary conditions~\cite{palassini} have
suggested otherwise and favor the droplet model.  None of these
studies has conclusively resolved which type of broken ergodicity
takes place in the low temperature phase in finite dimensions,
motivating a search for new approaches.

In this paper we propose and apply a rather direct method for determining the
number of PSPs for an Ising spin glass.  Our main ideas were introduced in a
previous Letter;~\cite{previous} here we provide an extended development of
these ideas, along with further analytical and numerical results.  Our method
involves the study of the spectral properties of the spin-spin correlation
function $\langle S_i S_j \rangle \equiv C_{ij}$. It is inspired by Yang's
analysis~\cite{Yang} of correlation functions to detect off-diagonal
long-range order (ODLRO) in superfluids. In the case of superfluids, Yang
showed that the existence of ODLRO is equivalent to the presence of an
extensive eigenvalue in the spectrum of the one-particle density matrix. We
argue that, in the case of classical spin glasses, the spectrum of the
spin-spin correlation function contains a distinct signature which allows one
to determine unambiguously whether or not there are many pure state pairs,
i.e. whether or not RSB occurs. First of all, it is clear that the presence
of at least one extensive eigenvalue signals the presence of long-range
order. What we further 
show is that the {\em number} of extensive
eigenvalues determines unambiguously the number of pure state    %103  %G1016 
pairs: the spectrum contains {\em exactly one} extensive eigenvalue if 
and only if
there is {\em exactly one} pure state pair, and it contains {\em more than
  one} extensive eigenvalues if and only if there are {\em more than one}
pure state pairs. We also argue that the extensive eigenvalues dominate the
trace of $C$, and that the nonextensive eigenvalues scale with the number $N$
of spins in the system with a power lower than $1$.  We confirm these
arguments by performing Exchange Monte Carlo \cite{Marinari,Hukushima} (EMC)
simulations for the SK model in various regimes, for which we know the nature
of the ergodicity breaking in the equilibrium state. That is, we find
multiple extensive eigenvalues in the spin-glass phase, but only a single
such eigenvalue in the ferromagnetic phase and in the paramagnetic phase in
a field.  In the RSB phase (where one expects many PSPs) we find that, for
the range of system sizes that we have studied, the eigenvalue spectrum is
dominated by a small number of extensive eigenvalues.  Making the simplifying
assumption that the thermal average is dominated by only {\it two\/} PSPs, we
are able to introduce an analytically calculable model which reproduces the
eigenvalue spectrum for the SK model surprisingly well.  Finally, we use EMC
simulations to study the Edwards-Anderson (near-neighbor) model in four
spatial dimensions.
Our results, for system sizes $ 3^4 \le N \le 6^4 $, are compatible with the
presence of {\em only one} extensive eigenvalue of $C$. The second largest
eigenvalue of $C$ is very well fit (with $\chi^2 = 0.132$) by a power-law scaling
with $N$, with an exponent (smaller than one) that is consistent with the
value predicted by the droplet theory.   %A103
An alternative fit with a form that allows for an extensive piece in the
second largest eigenvalue of $C$ is possible~\cite{Par_comm}, although it would imply that
the ratio of the eigenvalue over $N$ saturates to a finite value, something
that is not observed in our data.

We organize this paper as follows. In Sec.~\ref{SEC:ssc} we show how
the properties of the spectrum of $C_{ij}$ below $T_c$ can be used to
distinguish between RSB and a single PSP in the low temperature phase.
In Sec.~\ref{SEC:SK} we discuss numerical results for the SK model in its
various regimes. 
In Sec.~\ref{SEC:two_pure} we introduce a two-PSP model and show that it
gives results much like the numerical results for the SK model. 
In Sec.~\ref{SEC:EA4D} we present and discuss the results of our EMC
simulations for the short-range Edwards-Anderson model in four dimensions.
Finally, in Sec.~\ref{SEC:conclusions} we present our conclusions.

\section{Spin-spin correlators and pure states in Ising systems}
\label{SEC:ssc}
Guided by the analogy to Yang's analysis of the matrix elements of the
one particle density matrix in a superfluid system, we %% G1016 are going to
study the spectrum $\{ \lambda_{i} \}_{i=1,\cdots,N}$ of the spin-spin
correlation matrix $C_{ij}$ of an Ising spin system. This matrix has
an extensive trace $Tr{C} = \sum_{i} \langle S_{i}^{2} \rangle =
N$. In addition, it is positive semi-definite: for an arbitrary real  %A103
N-dimensional vector $|v\rangle$,
\begin{eqnarray*}
\langle v|C|v \rangle&=& \sum_{ij} v_i C_{ij} v_j = 
\sum_{ij} v_i \, \langle S_i S_j \rangle \, v_j \\
&=& \langle \Big( \sum_{i} v_i S_i \Big)^{2} \rangle \ge 0.
\end{eqnarray*}
Above the ordering temperature $C_{ij}$ decays exponentially to zero
at large $r_{ij}$, so it has no large eigenvalues. For $T\to\infty$
(or for $T>T_c$ in the SK model, for which distance is not
meaningful), the correlation matrix reduces to the identity
$C_{ij}=\delta_{ij}$, and consequently all its eigenvalues are equal
to one. Below $T_c$, however, $C_{ij}$ is nonzero almost everywhere
due to the ordering of the spins. For one PSP, in the low temperature
limit $T\rightarrow 0$, $C_{ij} \rightarrow \pm 1$, $C$ has one
eigenvalue which approaches $N$ as $T\rightarrow 0$, and the rest of
the eigenvalues go to zero.  Hence, there is a transition in the
distribution of the eigenvalues of $C_{ij}$ as $T$ crosses
$T_c$. Specifically, just as in the case of superfluids
and ferromagnets and antiferromagnets, we can detect       %A103
the existence of long range order below $T_c$ by the presence of one
(or more) extensive eigenvalue(s).  This approach based on examining
the spectrum of the correlation functions has a definite advantage in
the study of disordered systems: we can eliminate the necessity for
guessing the {\em nature} of the order---i.e., the eigenvectors
corresponding to the extensive eigenvalues---and still detect its {\it
existence\/}. We will also find that this spectrum can give unique and
clear information about the {\it number of pure states\/} in a frozen,
disordered phase.

Although the spin-spin correlation function $C_{ij}$ has not been
extensively used to probe the nature of the broken ergodicity in the
spin glass phase, many related quantities have been used to study some
of the static properties of spin glasses. For example, the quantity
\begin{equation}
q^{(2)} = [{\rm Tr}\,C^2/N^2]_{\rm av} 
\end{equation}
is a commonly used order parameter in spin-glass systems.~\cite{B&Y}
Also, the spin-glass susceptibility $\chi_{SG}^J $ for one disorder
realization (i.e. for a fixed set of spin couplings $J_{ij}$) 
\begin{equation}
\chi_{SG}^J = \frac{1}{N} \sum_{ij} (\langle S_i S_j \rangle - \langle
S_i \rangle \langle S_j \rangle)^2
\end{equation}
is given, for $T > T_{c}$, by
\begin{equation}
\chi_{SG}^J = \frac{1}{N} \sum_{ij} C_{ij}^2 = {\rm Tr}\,C^2/N.
\end{equation}
We see that the disorder average of $C^2$ contains interesting information
about the freezing of the spins (while in contrast the disorder average of
$C$ contains no information---it is just the unit matrix).  For $T<T_c$,
$q^{(2)}$ becomes of order $O(1)$, the largest eigenvalue $\lambda_{1}$ of
$C$ becomes of order $ N$, and ${\rm Tr}\,C^2 = \sum_i \lambda_i^2$ 
becomes of order $ N^2$. 
Hence our measure of order ($\lambda_{1}
\sim N$) is consistent with earlier measures used for spin glasses.

We wish further to obtain new and independent information from the
spectrum of $C$; towards this goal, in the following two subsections
we will argue that one can detect the presence or absence of order
with {\it many\/} valleys (i.e., RSB) simply by counting the number of
extensive eigenvalues of $C$.  To make this connection, the main idea
used will be that pure states are characterized by their {\em
clustering}~\cite{MPV} property, i.e. that the spin-spin correlation
function at long distances can be approximately decomposed as a linear
combination of (possibly non-orthogonal) projectors onto the subspaces
associated with the pure states present in the system, i.e.:
\begin{equation}
\langle S_i S_j \rangle \approx \sum_{\alpha} 
w^{\alpha} \langle S_i \rangle^{\alpha} \langle S_j \rangle^{\alpha}, 
\label{EQ:approx_decomp}
\end{equation} 
where $\langle \phantom{X} \rangle^{\alpha}$ denotes a thermal average restricted 
to the pure state $\alpha$. From this relation, a connection will be
obtained between the number of extensive eigenvalues of the left hand
side and the number of pure state pairs present in the right hand
side. The rest of this section is devoted to deriving this connection,
and to estimating the effects on the spectrum due to the terms
neglected in Eq.~(\ref{EQ:approx_decomp}). 

\subsection{Single pure state pair}
\label{SEC:1PSP}
We first consider the case of a single PSP. We show that in this case
there is only one extensive eigenvalue that dominates the trace of $C$.  

Without loss of generality, we can consider the case of only one pure state.
This applies directly if one of the pure states in the pair is selected by an
external field or a boundary condition. But even when the two pure states are
present, the spin-spin correlation matrix for the system is the same as if
only one of them was present, simply because it involves a product of an {\em
  even} number of spin variables.~\cite{FNOTE:one_state}

Hence we can rewrite $C_{ij}$ as
\begin{eqnarray}
C_{ij} & = & C^{0}_{ij} + V_{ij} \\
C^{0}_{ij} & \equiv & \langle S_i \rangle \langle S_j \rangle \\
V_{ij} & = & \langle S_i S_j\rangle_c, 
\end{eqnarray}
where $\langle S_i S_j\rangle_c \equiv \langle S_i S_j\rangle -
\langle S_i \rangle \langle S_j \rangle $ is the connected correlation
function. By arguments similar to the one used to show that $C$ is
positive semi-definite, both $C^{0}$ and $V$ are positive
semi-definite (see Appendix~\ref{app:M}). 

Let's ignore for the moment the connected part $V$, and concentrate on
the matrix $C^0$. This matrix is proportional to the projector onto
the vector $s_i = \langle S_i \rangle$, thus it has exactly one
extensive eigenvalue
\begin{equation} 
\kappa_{1} \equiv {\rm Tr} C^0 = \langle s|s\rangle =
\sum_i \langle S_i \rangle^2 = N q^{11} 
\end{equation} 
(where $q^{11}$ is the self-overlap of the pure state) with
eigenvector proportional to $s_i$, and $N-1$   %A103
eigenvectors with eigenvalue zero. 

We then ask how the matrix $V$ changes this distribution of
eigenvalues.  By the clustering property of pure states~\cite{MPV}, 
the typical element of $V$ vanishes as $N\to\infty$; hence it is reasonable
to assume that the typical value of the ratio between the largest eigenvalue
$\upsilon$ of $V$ and the largest eigenvalue $\kappa_{1}$ of $C^{0}$ goes to
zero in the thermodynamic limit. By a detailed variational argument (see
Appendix~\ref{app:strong_clustering}), it can be shown that the largest
eigenvalue $\lambda_{1}$ of $C$ is bounded between $\kappa_{1}$ and
$\kappa_{1} + \upsilon$, and in the case that $\upsilon \ll \kappa_{1}$ the
second largest eigenvalue $\lambda_{2}$ of $C$ is bounded above by $\upsilon$
times a number that goes to one. In other words, the largest eigenvalue of
the correlation matrix remains extensive (and, in fact, it changes very
little) when the effect of $V$ is included. This result can also be
recovered more intuitively by applying perturbation theory to the problem of
estimating the effect of $V$.

The problem that remains to be solved is, therefore, the estimation of
the largest eigenvalue of $V$.  A possible assumption would be that
the off-diagonal parts of $V$ have a typical behavior in the large-$N$
limit of $V_{ij} \sim N^{-\delta}$, with $\delta >0$.  (The diagonal
elements are always $\sim N^{0}$, but they have an effect of order
$ N^{0}$ on the eigenvalues.) The largest eigenvalue of $V$, and   %G1016 
therefore the second largest eigenvalue of $C$, are then of order
$N^{1-\delta}$. One can also view this result in the following
way. $V$ can at most reweight the eigenvalues of $C$ as if it is an
additional pure state with thermodynamic weight $\propto N^{-\delta}$
(see Sec.~\ref{SEC:two_pure}), in which case it gives rise to an eigenvalue (again)
$\propto N^{1-\delta}$.  An eigenvalue of order $N^{1-\delta}$ is, in
principle, distinguishable from an eigenvalue of order $N$, since $\delta >
0$.

In some cases it is possible to obtain a stronger bound on the decay
exponent $\delta$.  For instance, if we assume that $\chi_{SG} \sim
N^{0}$, then we get $\delta \ge 1/2$ by simple power counting.  This
should be the case everywhere above the Almeida-Thouless (AT)
line~\cite{AT} in the phase diagram~\cite{phasediag} for the SK model.
Above this line there is a paramagnetic phase with a single pure
state, with long-range order trivially induced when the external
magnetic field $h$ is nonzero.  There is also a conventional
ferromagnetic phase with a single PSP, in another part of the phase
diagram but still above the AT instability.  Either of these phases
should have $V_{ij}$ decaying with $\delta \ge 1/2$.  For a pure
ferromagnet with no frustration or disorder, assuming (as appropriate
for a single PSP) the uniform susceptibility $\chi = (1/N)\sum_{ij}
\langle S_i S_j\rangle_c$ is of $O(1)$ gives the stronger constraint
$\delta = 1$.  We expect this latter limit to be approached for the SK
problem, when either the external magnetic field $h$ or the average
ferromagnetic coupling $J_0$ are sufficiently large.

While the above arguments make no direct reference to spatial
dimension, they do rely on the notion of a `typical' element of
$V$. This idea is certainly appropriate for infinite-range models such
as the SK problem; and the above arguments may also be applied to any
finite-dimensional problem for which a typical behavior of $V_{ij}$
can be defined.  For example, one can define a typical $V_{ij} =
\langle S_i S_j \rangle_c$ for any magnetic system with a finite
correlation length; here the typical $V_{ij}$ is exponentially small
and so the matrix $V$ does not have significant effects on the
eigenvalues of $C^0$.  However, correlations in finite-dimensional
spin glasses, in the frozen phase, are thought to fall off more slowly
than exponentially, \cite{Bokil} giving an infinite spin-glass
susceptibility in the spin-glass phase~\cite{droplet,FH} and rendering
the notion of a typical element of $V$ problematic.  

Hence we examine carefully the `droplet' theory,~\cite{droplet} which
is the outstanding candidate for a theory of finite-dimensional spin
glasses without RSB.  In this theory the low energy excitations at
large distances are assumed to be large droplets of collectively
overturned spins of size $L$, whose energy scales as $L^\theta$.  It
follows that the majority of the elements of $V_{ij}$ are
exponentially small; however there is also a set of `big' elements
which are of $O(1)$ in magnitude. These elements occur when $i$ and
$j$ lie within the same `active' (coherently flipping) droplet; this
makes $\langle S_i \rangle$ and $\langle S_j \rangle$ small, while
leaving $\langle S_i S_j \rangle$ and $\langle S_i S_j \rangle_c$
large. The fraction of these big elements is of order $ 1/L^\theta$, where  %%G1016 
$L$ is the system size and $\theta$ is a scaling exponent from the
zero-temperature fixed point. Although this `big' fraction vanishes in
the thermodynamic limit, it still can have large effects at large $N$.
For example, let us suppose that a finite sample of size $L$ is
dominated by one large active droplet of size of order $L$. In such a
case the big elements of $V$, appearing with probability $\sim
1/L^\theta = 1/N^{\theta/d}$, are coherent, so that the largest
eigenvalue of $V$ is of order $N\times N^{-\theta/d} = N^{1-\theta/d}$. (We
have verified this with simple numerical experiments.)  Given this
bound on the eigenvalues of $V$, the largest eigenvalue of $V$ is
again much smaller than the largest eigenvalue of $C^0$, and again the
second largest eigenvalue $\lambda_{2}$ of $C = C^0 +V$ is of order
$N^{1-\theta/d}$ or smaller. Hence one gets a decay exponent
$\delta_{2}$ for $\lambda_{2}/N$ equal to $\theta/d$. Given the
assumptions leading to this conclusion, this value is a lower bound
for the rate of decay of $\lambda_{2}/N$. It is plausible however that,
for a range of system size $N=L^d$ not too large, the assumption of
dominance by a single active droplet can hold, giving this lower bound
for $\delta_{2}$. Furthermore, the numerical value of the latter can be
quite small: about $0.19/3 \approx 0.063$ in three dimensions,
\cite{BM84} and about $0.7/4 \approx 0.17$ in four.~\cite{RBY90} Hence one
needs good numerical data for $\lambda_{2}(N)$ in a finite-dimensional
spin glass to distinguish RSB (with $\delta_{2}=0$) from a $\lambda_{2}/N$
which is weakly decaying due to droplet excitations.

\subsection{Many pure state pairs}

In the previous subsection we have presented strong arguments,
assuming there is only a single PSP in the low temperature phase, that
there can be only one extensive eigenvalue of $C$. It follows from our
argument that the observation of more than one extensive eigenvalue
directly implies RSB. In short, letting $N_{PSP}$ be the number of
PSPs, and $N_{EEV}$ be the number of extensive eigenvalues, we found
\begin{equation}
(N_{PSP} = 1) \Longrightarrow (N_{EEV} = 1) 
\label{forward}
\end{equation}
or
\begin{equation}
(N_{EEV}>1) \Longrightarrow (N_{PSP} > 1) \equiv ({\rm RSB}).
\label{not}
\end{equation}
Now we would like to argue that the converse is also valid, i.e., to
find some necessary consequence of RSB in the eigenvalue spectrum.
Hence we will assume RSB (that there is more than one pure state 
%% HC1019
pair
present), and then determine how many extensive eigenvalues there should  %%G1016 
be in the spectrum of the spin-spin correlation matrix.

Let us suppose that there are $p \ge 1$ pure states, characterized by
the magnetizations $ \{ m_i^\alpha \}$ ($m_i^\alpha \equiv \langle
S_i\rangle^\alpha$), and thermodynamic weights
$\{w^\alpha\}$, with $\alpha=1,\cdots,p$. Here we only include pure
states whose thermodynamic weight $w^\alpha$ is nonvanishing. We now
decompose the correlation function into the contributions coming from
each pure state:
\begin{eqnarray}
C_{ij} & = & \sum_{\alpha=1}^p w^\alpha \langle S_i S_j \rangle^\alpha\\
 & = & C^0_{ij}+V_{ij}\,\,, 
\end{eqnarray}
with
\begin{eqnarray}
C^0_{ij} & = & \sum_{\alpha=1}^p w^\alpha m_i^\alpha m_j^\alpha 
\label{EQ:C0_def} \\
V_{ij} & = & \sum_{\alpha=1}^p w^\alpha V_{ij}^\alpha \\
V_{ij}^\alpha & = & \langle S_i S_j \rangle_{c}^\alpha. 
\end{eqnarray}
By the clustering property of pure states, we may assume that $V$ is
small compared with $C^0$. As in the case of one PSP, we will proceed
in two steps. First we will study the spectrum of $C^0$, and later we
will include the effect of $V$.

Let's define the $p$ vectors $ \{ | \phi_{r} \rangle \}_{r=1,\cdots,p} $
\begin{equation}
(\phi_{r})_j=\sum_{\beta=1}^p c_\beta^{r}
\sqrt{w^\beta} m^\beta_j,
\label{proposed}
\end{equation}
with the coefficients $c_\beta^{r}$ to be determined later. For an
appropriate choice of the $c_\beta^{r}$, these
vectors can be shown to be eigenvectors of $C^0_{ij}$. In fact, 
\begin{eqnarray}
\sum_{j=1}^N C^0_{ij} (\phi_{r})_j
&=&\sum_{j=1}^N \sum_{\alpha,\beta=1}^p
w^\alpha m_i^\alpha m_j^\alpha c_\beta^{r} \sqrt{w^\beta} m^\beta_j
\nonumber\\
&=& N \sum_{\alpha=1}^p \sqrt{w^\alpha} m_i^\alpha \sum_{\beta=1}^p
A_{\alpha \beta} c_\beta^{r}\,\,,
\end{eqnarray}
where we have defined the real, symmetric $p\times p$ matrix
$A_{\alpha \beta} \equiv \sqrt{w^\alpha} q^{\alpha \beta} \sqrt{w^\beta}$.
The matrix $A_{\alpha \beta}$ has $p$ orthonormal eigenvectors $\{
c_\beta^{r}\}$ with eigenvalues $\{{a}_{r}\}_{r=1,\cdots,p}$.
By inserting one of these eigenvectors in Eq.~(\ref{proposed}) we
obtain
\begin{equation}
\sum_{j=1}^N C^0_{ij} (\phi_{r})_j=N {a}_{r} (\phi_{r})_i\,\,.
\label{C^0}
\end{equation}
Thus, for each nonzero eigenvalue ${a}_{r}$ of $A$, an eigenvector
of $C^0$ is obtained with eigenvalue $\kappa_{r} = N {a}_{r}$. Let's
denote by $\bar{p}$ the number of linearly independent magnetization
vectors among the magnetizations of the $p$ pure states ($1\le \bar{p}
\le p$).  This number need not be equal to $p$: for example, the
magnetizations of the two pure states in a PSP are not linearly
independent, since one of them is $-1$ times the other. Having
$\bar{p} = 1$ is equivalent to saying that there is either only one
pure state present, or there is exactly one PSP. Therefore, for the
droplet picture $\bar{p}=1$ and for the RSB picture $\bar{p}> 1$
strictly, i.e.:
\begin{equation}
(N_{PSP}>1) \Longrightarrow (\bar{p} > 1).
\label{EQ:rsb_pg1}
\end{equation}
It is an exercise in linear algebra to show that (see
Appendix~\ref{app:M}): (i) the number of nonzero eigenvalues of $A$  %%G1016
is exactly $\bar{p}$, (ii) all of these nonzero eigenvalues are  %%G1016
positive, (iii) the corresponding eigenvectors of $C^0$ are linearly  %%G1016
independent, and (iv) the remaining $N-\bar{p}$ linearly independent  %%G1016
eigenvectors of $C^0$ have zero eigenvalue and are orthogonal to all
pure state magnetizations.  As a consequence, the number of nonzero 
extensive eigenvalues of $C^0$ is equal to $\bar p$, the number of pure
states with linearly independent magnetizations.

Next we assess the effects of $V_{ij} = \sum_{\alpha} w^\alpha
V_{ij}^\alpha$.  We assume that the sum over pure states is finite
(see below, and Ref.\ \onlinecite{MPSTV}).  Hence, even if the
$V_{ij}^\alpha$ decay at different rates with $N$, we can still take
the typical element of $V_{ij}$ to decay with $N$ at least as fast as
$N^{-\delta}$ for some $\delta>0$. Hence the largest eigenvalue of $V$
is of order $N^{1-\delta}$ or smaller. From this we can show (Appendix~\ref{app:strong_clustering})
that there are still $\bar p$ extensive eigenvalues for $C = C^0 + V$,
\begin{equation}
N_{EEV} \ge \bar{p}.
\label{EQ:neev_p}
\end{equation}
Combining this with Eq.~(\ref{EQ:rsb_pg1}) it follows that
\begin{equation}
(N_{PSP}>1) \Longrightarrow (N_{EEV} > 1). \ \;
\end{equation}
that is, the converse of Eqs.\ (\ref{forward}) and (\ref{not}).

It is plausible, although {\em not} proven, that in general a complete
set of pure states not related by spin inversion (thus constituting
one-half of the total set of pure states) will all be, with
probability one, linearly independent, so that $\bar p$ is just the
number of PSPs, i.e. $N_{PSP} = N_{EEV}$.

Note that, in the Parisi RSB solution~\cite{MPV} to the mean-field
problem, the number of PSPs grows with $N$, at a rate which is not known.
The ultrametric structure of these pure states implies~\cite{Baldi}
that they cannot grow in number faster than $N$. However there is a stronger
constraint, coming from Eq.~(47) of Ref.\ \onlinecite{MPSTV}, which states that
$\sum_\alpha (w^\alpha)^2 = O(1)$. This tells us that the diverging number
of pure states do not have equal thermodynamic weight; instead a finite 
number of them dominate the sum of the weights $w^\alpha$, with the
rest having negligible weight. 

One might ask whether the ultrametric structure of the space of pure states
might imply some constraint on the number which are linearly independent.
However we find no such constraint in general. For example, one can construct,
for any number $n \le N$ (where $N$ is the dimension of the vector space),
a set of $n$ vectors which are both linearly independent and ultrametric;
but one can also construct a set which is linearly dependent and ultrametric.

It may be possible to derive tighter
bounds on these quantities via further theoretical work. In the following
section we provide some further information, obtained from equilibrium Monte-Carlo
studies, on $N_{EEV}$ for a finite range of $N$ in the SK problem.

\section{Results for the SK model}
\label{SEC:SK}
The SK model is described by the Hamiltonian 
\begin{equation}
{\cal H}=-\frac{1}{2}\sum_{i\ne j} J_{ij}S_i S_j -h\sum_i S_i\,\,,
\label{Ham}
\end{equation}
%%G1016 
with $[J_{ij}]_{\rm av} = J_0/(N-1)$ and $[J_{ij}^2]_{\rm av} -
[J_{ij}]_{\rm av}^2 = J^2/(N-1)$ for any $i$ and $j$.  This model,
which is equivalent to an infinite dimensional model for which mean
field theory is exact, has a phase diagram~\cite{phasediag} in
($h/J$,$J_0/J$,$T/J$) space which is reasonably well understood. In  %A103
particular, there are instability lines (which presumably form a
surface) below which the replica symmetric    %A103
solution is unstable, and the Parisi RSB
ansatz~\cite{Parisi} is believed to give the correct solution.  The
`AT line' found by de Almeida and Thouless~\cite {AT} lies in the
$h$-$T$ plane (i.e., $J_0 = 0$); below this AT line there is RSB, while
increasing either $h$ or $T$ brings one to a phase consisting of a
single pure state. This phase is continuous with the paramagnetic
phase at $h=0$; it has long-range order which is trivially induced by
the field, and hence neither spontaneous symmetry breaking nor broken
ergodicity of any other sort. Nevertheless we expect a large
eigenvalue for $C$ due to the long-range order.  In another region of
parameter space ($h=0$, with a ferromagnetic bias $J_0$ sufficiently
large) there is a ferromagnetic phase with one PSP.  Here one has the
familiar version of broken ergodicity in the form of spontaneous
symmetry breaking; given that there is a single PSP, we expect a
single extensive eigenvalue in this phase also.  Thus we find three
distinct phases (glass/RSB, paramagnet+field, ferromagnet) which we
can explore via Monte Carlo simulations in order to test our ideas
about correlation functions, eigenvalue counting, and ergodicity
breaking in Ising systems.

We have performed EMC~\cite{Hukushima,Marinari} simulations for the SK
model at three points in the phase diagram.  In glassy systems with
very long relaxation times, normal Monte Carlo simulations are limited
to small system sizes because of the divergent relaxation and
equilibration times.~\cite{Young85} EMC simulations allow for the
crossing of barriers in a reasonable simulation time, via a stochastic
walk of each simulated system not only in configuration space but also
in temperature.  The simulation consists of having many systems at  %%G1016 
different temperatures for the same disorder realization running in
unison, and attempting to exchange the configurations between adjacent
temperatures after a given number of Monte Carlo steps $\delta t$. The
exchange of neighboring temperature configurations takes place with
the probability
\begin{equation}
P(S_m\leftrightarrow S_{m+1};\beta_m,\beta_{m+1})=e^{-\Delta}\,,
\end{equation}
with
\begin{equation}
\Delta\equiv (\beta_m-\beta_{m+1})(E(S_{m+1})-E(S_m))\,,
\end{equation}
and $S_m$ indicating the instantaneous spin configuration at
temperature $1/\beta_{m}$. With this probability of exchange one
ensures that the systems at the different temperatures remain in
thermal equilibrium---whether or not they exchange temperatures---at
all times during the simulation. This is because the configuration
obtained from an exchange with a higher temperature is still accepted
with the normal Boltzmann probability for the lower (accepting)
temperature; hence an exchange drives neither system out of
equilibrium.  If the temperature difference between neighboring
systems is not too large, then exchanges are accepted at a reasonable
rate, and each system explores the full range of temperatures.  Hence
each system is effectively cooled and heated many times during the
simulation, ensuring that all barriers have been effectively
circumvented, not by crossing them but by falling within their
boundaries from a higher temperature.  An important parameter in this
simulation is the spacing between the different temperatures, which
must be adjusted to get a high enough acceptance ratio for the
temperature exchanges.~\cite{Marinari,Hukushima}

Our criteria for having reached thermal equilibrium in our measured
quantities involve calculating the spin-glass susceptibility for a
single disorder realization using two distinct methods. One method,
discussed extensively in previous studies,~\cite{Bhatt&Young87} uses
the averaging of the overlap of two uncoupled replicas
\begin{equation}
\chi_{SG}^{(1)}=
\frac{1}{N}\frac{1}{\tau_0}
\sum_{t=1}^{\tau_0}
\left [ (\sum_i S_i^{(1)}(t_0+t)S_i^{(2)}(t_0+t))^2\right ],
\end{equation}
and the other uses the standard way of calculating the thermal average
in Monte Carlo simulations.
\begin{equation}
\chi_{SG}^{(2)}=
\frac{1}{N}\sum_{ij}\left\{\frac{1}{\tau_0}
\sum_{t=1}^{\tau_0} S_i(t_0+t)S_j(t_0+t)\right\}^2\,.
\end{equation}
Here $\tau_0$ and $t_0$ have to be chosen large enough to obtain
thermally equilibrated results. We also demand full symmetry of the
overlap distribution function $P(q)$, and that all initial
configurations visit all temperatures evenly.  We have also checked
our results using commonly studied quantities such as $P(q)$ and
$\chi_{SG}$ doing standard Monte Carlo simulations for the smaller
system sizes; here our results are in agreement with previous
work.~\cite{Bhatt&Young87,Young85} In our simulations we have used a
temperature spacing of $\Delta T=0.05 J$, $t_0=3-5 \times 10^4$,
$\tau_0=10-40 \times 10^4$, and $\delta t=10-20$.  (Here all times are in
units of Monte Carlo steps per spin.)

We find that the eigenvalue spectrum of $C$ is strongly dependent on
disorder realization, such that it is impossible to observe any
regular dependence of the $\lambda_i$ on $N$ with increasing $N$ for a
single set of $J_{ij}$'s. (See Fig.\ \ref{CSD} below for examples of
similar behavior in four space dimensions.)
Hence it is necessary to accumulate statistics on the eigenvalue
spectra for many disorder realizations.  The eigenvalue probability
distribution for the first two eigenvalues of $C_{ij}$ for $N=128$ at
zero field and at $h=1.2 J$ (above the AT line)---all at $T=0.4
J$---are shown in Fig. \ref{distrib}. These distributions have been
obtained from 3400 disorder realizations. %%JAIRO: is this true? HORACIO:Yeap
It is clear from Fig. \ref{distrib} that, at least in the RSB phase, the
distributions for $\lambda_{1}$ and $\lambda_{2}$ are extremely broad;
also they show significant skew. Hence we have studied both $[
\lambda_i ]_{\rm av}$ and $[\lambda_i]_{\rm typ}\equiv\exp[\ln
\lambda_i]_{\rm av}$ for small $i$ in each phase.

We show the average of the ten largest eigenvalues as a function of
system size in Fig.\ \ref{SK_ave} for different points of the SK phase
diagram: (a) $h=0$, $T/J=0.4$, $J_0 =0$, the RSB phase; (b) $h/J =
1.2$, $T/J=0.4$, $J_0 =0$, the paramagnet in field with one pure
state; and (c) $h=0$, $T/J=0.4$, $J_0/J = 1.5$, the ferromagnet with a
single PSP.
The system sizes considered are $N=32,\, 64,\, 128,\, 192,\,256,\,$
and $512$ with $3400,\, 3400,\, 3400, \, 1400,\, 1100,\,$ and $400$
disorder realizations performed respectively.  It is clear from Fig.\
\ref{SK_ave}(a) that two eigenvalues are of order $N$ for $N \agt
200$. This is extremely strong~\cite{fse} numerical evidence for more than one PSP, and %%G1016 
hence nontrivial ergodicity breaking.  We expect further $O(N)$
eigenvalues to emerge for larger $N$, as suggested by the behavior of
$[ \lambda_3 ]_{\rm av}$ in the figure.

In contrast, there is only one large eigenvalue in Fig.\ \ref{SK_ave}(b)
and (c).  We find further that $[ \lambda_{2}]_{\rm av}/N $ may be fit to
a power law $N^{-\delta_{2}}$ for some range of $N$ in cases (b) and (c).
In the paramagnetic phase (b) $\delta_{2} \sim 0.52$ while in the 
ferromagnet (c) $\delta_{2}$ is somewhat larger, $\sim 0.58$.
While we do not expect these power laws to have reached their asymptotic
values for the system sizes considered here,   %A103 
we do expect any observed decay for large enough $N$ to be
consistent with our arguments in the previous section, where we obtained
$\delta_{2}\ge 1/2$ in this regime. MC results for the paramagnetic phase,
in a larger field than in (b), show that
$[ \lambda_{2}]_{\rm av}/N$ decays with a larger exponent (we have observed up to
$\sim 0.75$), which we expect to approach 1 for large enough $h$.
We also show in Fig. \ref{SK_typ} the scaling of the typical value
$[\lambda_i]_{\rm typ}/N$ for the first ten eigenvalues in the RSB phase.
Here the behavior is qualitatively similar to that of 
$[ \lambda_i]_{\rm av}/N$ in Fig.\ \ref{SK_ave}(a): both figures show
clearly that $[ \lambda_{2}]/N$ (av or typ) is flat as a function of $N$  %%G1016 
above a threshold value for $N$ which is of order 100--200, plus
strong signs that $[ \lambda_3]/N$ and $[ \lambda_4]/N$  %%G1016 
are also approaching a flat behavior. Hence we see graphically, in these
figures, the emergence of multiple PSPs with increasing $N$.

Figs.\ \ref{SK_ave} and \ref{SK_typ}, taken together,
give convincing evidence that the eigenvalue spectrum of $C$ 
can clearly distinguish trivial (one PSP) from nontrivial broken ergodicity. 
This spectrum allows one, for large enough $N$, to detect multiple PSPs 
simply by counting the number of extensive eigenvalues $N_{EEV}$.   %A103
As discussed above, while the number of PSPs is believed to diverge in the 
thermodynamic limit, 
%%% HC1019
only a few of them dominate the thermodynamics;~\cite{MPSTV} 
we see some of these emerging
in Figs.\ \ref{SK_ave}(a) and \ref{SK_typ}. 
%%% HC1019
A possible consequence of this would be that, for large enough $N$, $N_{EEV}$
might saturate at a finite constant.
The present data show no clear sign of this, although they 
are clearly beyond the threshold of $N$ for which $N_{EEV}$ begins to
exceed $1$. For $N$ in the vicinity of this
threshold, we expect that one can reasonably view the glass phase
as having just {\it two\/} PSPs; and we explore the consequences of 
this assumption next.

\section{Two pure state pairs}
\label{SEC:two_pure}
As suggested by Figs.\ \ref{distrib}, \ref{SK_ave}, and  \ref{SK_typ},
the spectrum of $C$ for the SK problem is dominated by two
large eigenvalues, each coming from a broad distribution of values, for the system
sizes considered here. We can understand much of this behavior with 
a simple two-PSP model which can be computed analytically.
We begin with a very simple model for two PSPs at $T=0$.
Suppose that phase space consists of only two spin configurations 1 and 2, 
and ignore all others.
Take $C(w)= w C_{1}^0+(1-w)C_{2}^0$, with $C_{1}^0$ and $C_{2}^0$ 
corresponding to the $C_{ij}$ of the two different configurations 
$S^{0}_{1i}$ and $S_{2i}^{0}$ at zero temperature, and $w$ (a thermodynamic
weight) ranging from 0 to 1/2. The overlap between the two states 
is given by $q_{12}=\sum_i S^{0}_{1i} S_{2i}^{0}/N$.
It can be easily shown that this matrix has only two
nonzero eigenvalues, corresponding to  
\begin{equation}
\frac{\lambda_\pm (q_{12},w)}{N}
=\frac{1\pm \sqrt{q_{12}^2+(1-q_{12}^2)(2w-1)^2}}{2}\,\,.
\label{lambdapm}
\end{equation}
Note that $\lambda_+$ ranges from $N$ at $w=0$ to 
%%% HC1019
$(1+|q_{12}|)N/2$ 
%%% $(1+|q_{12}|)/2$
at $w=1/2$. 
It is also clear that, for small $w$, $\lambda_-$ is linear in $w$;
hence if $w$ is of lower order in $N$ than $N^0$, $\lambda_-$ ceases to be
extensive.
At the same time, even if the two PSPs have equal thermodynamic weight,
the second eigenvalue $\lambda_-$ approaches zero as $|q_{12}|\to 1$
and $N/2\,\,$ as $|q_{12}| \to 0$.
This is consistent with our results from Section~\ref{SEC:ssc}, since as
$|q_{12}|\to 1$ the two states become linearly dependent and $\bar p$ becomes $1$.
We can proceed further with this two-PSP model by
calculating the probability distribution of $\lambda_{1}$ and $\lambda_{2}$, as follows:
\begin{eqnarray}
&&\tilde{P}_w(\lambda_{1})=\int d\,q \,\,P_{12}(q) 
\delta(\lambda_{1}-\lambda_+(q,w))
\nonumber\\
&&=\frac{2\sqrt{q_0^2+(1-q_0)^2(2w-1)^2}}{q_0(1-(2w-1)^2)}
\theta(\lambda_{1}-\lambda_{+min})
P_{12}(q_0)\,\,,
\end{eqnarray}
where $P_{12}(q)$ is the probability distribution of $q_{12}$,
$\lambda_{+min}$ depends on $w$, and $q_0$ is an implicit function of
$\lambda_{1}$ given by $\lambda_+(q_0,w)=\lambda_{1}$. 
For the case of $w=1/2$ this simplifies to
\begin{equation}
\tilde{P}_{w=1/2}(\lambda_{1})=2 P_{12}(2\lambda_{1}/N-1)
\theta(\lambda_{1}-\lambda_{+min})\,,
\end{equation}
and
\begin{equation}
\tilde{P}_{w=1/2}(\lambda_{2})=2 P_{12}(1-2\lambda_{2}/N)
\theta(\lambda_{-max}-\lambda_{2})\,;
\end{equation}
here $\lambda_{+min} = \lambda_{-max} = 1/2$.
At this simple level of approximation we already have the first indication
that the probability distribution of the first and second eigenvalues will be
very broad in the case of non-trivial broken ergodicity---as we have
seen in the MC results. 
Note however that in spite of the breadth of the distribution
$[\lambda_\pm ]_{\rm av}$ are still proportional to $N$.
Let us now augment this picture with finite-temperature effects. 
An approximate way to introduce temperature into the two-PSP model is 
as follows. We let
\begin{equation}
C_{ij}^{(1),(2)}=\left\{
\begin{array}{cc}
1 & \,\,{\rm for} \,i=j\\
\sigma_i^{(1),(2)} \sigma_j^{(1),(2)}\,&\, {\rm for}\,\,i\ne j\,\,,
\end{array}
\right.
\end{equation}
where $\sigma_i^{(1)} = \langle S_i \rangle^{(1)}$ and 
$\sigma_i^{(2)} = \langle S_i \rangle^{(2)}$ 
are gaussian random variables with the mean of
$\sigma_i^2$ equal to $q_{max}$. We adjust the distribution of
$\sigma_i^{(1)}$ and $\sigma_i^{(2)}$ such that 
$q_{max}$ agrees with that obtained from our MC runs, while
also demanding that the overlap distribution 
$(1/N)\sum_i \sigma_i^{(1)} \sigma_i^{(2)}$ 
agree with $P_{12}(q)$---which can also be extracted from our MC $P(q)$,
by subtracting a Gaussian part due to the self-overlap $P_{11}(q) = P_{22}(q)$.
The result of this decomposition procedure is illustrated in Fig. \ref{overlap}. 
This gives us a means of generating a realistic sample of $C$ matrices
corresponding to two PSPs.
We can then obtain the first few eigenvalues of 
$C_{ij}(w)=w C^{(1)}_{ij}+(1-w)C^{(2)}_{ij}$ and compare
them with the eigenvalues obtained directly from the MC simulations of the SK model. 
The former eigenvalues may be obtained either by an approximate analytical
perturbation approach (using only the diagonal part of $V$),
or by direct numerical
diagonalization of the matrix $C(w)$---which must in any event 
be generated by a random number generator.
Fig. \ref{nicefit} shows the eigenvalue distributions obtained from our
two-PSP model with $w = 1/2$, compared with those obtained directly from 
the EMC runs, for the SK model in the glass phase at $N=64$.
It is encouraging that our simple picture of two PSPs, with a minimal
set of assumptions, reproduces both the position and the shape of the two
distributions. 

We believe---and Fig.\ \ref{nicefit} supports this belief---that the
assumption of two PSPs has validity for a range of $N$ near the
threshold where RSB first appears.  We also believe that this
assumption will fail for larger $N$; our Monte-Carlo results strongly
suggest that the number $\bar p$ of significant PSPs will exceed two
as $N$ grows. It is interesting to ask what the large-$N$ limit of
$\bar p$ is. We obtain no answer to this question from the
considerations of Section~\ref{SEC:ssc}; while our numerical results suggest only
that $\bar p$ is at least as large as 3 or 4. We note here that Fig.\
\ref{nicefit} itself may be viewed as giving some indication of a
third large eigenvalue, if we assume (as is plausible from our two-PSP
results) that the third PSP robs weight from the upper part of
$\tilde{P}(\lambda_{2})$.

\section{Results for the EA model in four dimensions} 
\label{SEC:EA4D}
We have performed EMC studies of the four-dimensional Ising spin glass
on a hypercubic lattice,
with nearest-neighbor interactions, periodic boundary conditions,
and a gaussian distribution of the $J_{ij}$'s with zero mean.
The parameters of the calculation follow closely those of
Ref. \onlinecite{Marinari}.  We have focused on the point $T=J$ and
$h=0$.  These simulations are rather deep in the frozen phase, since
$T_c \approx 1.75 J$.  We choose this low temperature in order to try
to avoid spurious effects from closeness to the critical region; such
effects are likely to make it difficult to distinguish multiple PSPs
from a single PSP. The price we pay is that our MC runs are slow to
converge, while---according to the estimates of Ref.\
\onlinecite{Bokil}---we are still not fully out of the critical
region.  Our criteria for convergence are the same here as those we
used for the SK model (Sec.~\ref{SEC:SK}).

We have only examined the frozen phase for this problem. We find that
a plot of the distributions for $\lambda_{1}$ and $\lambda_{2}$ gives
broad and skewed forms similar to those seen for the RSB phase in
Fig.\ \ref{distrib}.  To complement the picture given in Fig.\
\ref{distrib}, we show in Fig.\ \ref{CSD} some examples 
%%% HC1019
of
%%%
the typical
behavior of a single disorder realization at ``fixed" $J_{ij}$'s and
increasing $N$. Here ``fixed" is in quotation marks since (as is well
known for spin glasses) adding spins requires adding bonds, and hence
a change in the $\{ J_{ij} \}$, which can often have nontrivial
effects.  Fig.\ \ref{CSD} bears out this expectation: the eigenvalue
spectrum of $C$ shows a highly irregular behavior as a function of
$N$.  If this irregularity were to persist in the limit $N\to\infty$
(such that the eigenvalues, and hence the correlations, had no
well-defined limit), then according to Newman and Stein,~\cite{csd}
there must be more than a single pure-state pair.  That is, ``chaotic
size dependence" is believed to characterize glasses with RSB, but not
to occur for a single PSP (unlike chaotic temperature
dependence~\cite{Banavar}). While the behavior shown in Fig.\
\ref{CSD} is interesting in this regard, we do not believe any
conclusion can be drawn from these data due to the small size of 
the systems considered here. %J925 
Instead we will focus
on trying to count extensive eigenvalues---a strategy that worked well
for the SK problem.  Fig.\ \ref{CSD} suggests rather strongly that
$\lambda_{1}$ is extensive; but it is impossible to draw any conclusion
about $\lambda_{2}$, and so we again resort to disorder averaging.

In Fig. \ref{EA4D} we show (a) $[ \lambda_i ]_{\rm av}/N$ and (b)
$[\lambda_i]_{\rm typ}/N$ for the first ten eigenvalues of $C$, at
$T=J$ and $h=0$.  The system sizes shown in Fig. \ref{EA4D} are
$N=81$, $256$, $625$, and $1256$ with $4000$, $1600$, $1300$, and
$400$ disorder realizations respectively.  These data suggest
that $[\lambda_{2}]_{\rm av}/N$ and $[\lambda_{2}]_{\rm typ}/N$ are each
decaying with $N$, with a clean power law $N^{-\delta_{2}}$.
A fit of the data in Fig. \ref{EA4D} gives (a) for the average eigenvalues,
$\delta^{\rm av}_{2} \sim 0.11$ and (b) for the typical eigenvalues,
$\delta^{\rm typ}_{2} \sim 0.15$.
The exponent we find for $[\lambda_{2}]_{\rm typ}/N$ is close to that
expected from our argument of Section~\ref{SEC:1PSP}, coupled with previous estimates
for the exponent $\theta$.  The latter range~\cite{Hartmann,Hukushima} from
0.6 to 0.8, while a simple extrapolation~\cite{RBY90} suggests $\theta \sim
0.7$. Our own argument ($\delta_{2} \ge \theta/4$) predicts a minimum value
for $\delta_{2}$ in the range 0.15 to 0.2; hence the behavior of
$[\lambda_{2}]_{\rm typ}/N$ is roughly consistent with this prediction, while
$\delta^{\rm av}_{2}$ for $[\lambda_{2}]_{\rm av}/N$ is somewhat smaller.

The decay of $[\lambda_{2}]/N$ is quite regular.  %%G1016 
Moreover, there is rough agreement between the exponent obtained from
the typical eigenvalues and the exponent estimated by assuming the
droplet picture to be valid. This evidence seems to favor the scenario
of only one PSP being present. 

Of course, it is in general difficult to settle from numerical
simulations alone any question that involves behaviors of a physical
system in the thermodynamic limit, since the results observed for some
sizes can change when larger sizes are simulated. In particular, %%G1016 
any claim that RSB does not occur, based on our method, %%G1016 
is necessarily more tentative than a conclusion that it does occur.~\cite{fse} %%G1016 
Thus %%G1016 
the results we obtain for $N \le 1296$ cannot rule out the possibility
of further PSPs appearing at some larger $N$, indicating 
%%that the RSB picture is correct. 
that there is RSB in the spin-glass phase in four dimensions.  %%G1016 
By contrast, if the droplet picture is correct for
4D spin glasses, then we believe that the decay exponent $\delta_{2}$
should, for sufficiently large $N$, increase from its lower bound as
the dominance of a single droplet fails.  Hence any sign of curvature
of the log-log plot of $\lambda_{2}/N$ vs.\ $N$, in either direction,
would be of significant interest.

It is also of interest to push results like those of Fig.\
\ref{CSD} to larger $N$. Here one seeks signs of convergence (or lack
of it) to a limit.  This criterion is, we believe, more difficult to
assess than the criterion we have applied to Figs.\ \ref{SK_ave},
\ref{SK_typ}, and \ref{EA4D}.  The latter criterion has the nice
property that one must only ascertain whether an integer---the number
of extensive eigenvalues---is one, or greater than one. However
studies seeking chaotic size dependence can certainly complement
studies of disorder-averaged eigenvalue scaling. 

\section{Conclusions}
\label{SEC:conclusions}
In this work we have applied the old idea of studying the eigenvalue spectrum of %%G1016 
a correlation function---used by Yang~\cite{Yang} to characterize  %%G1016 
ODLRO in superfluids---to  %A103  %%G1016 
a decades-old question in spin-glass physics, namely: how many pure
states are there in the frozen phase, and how are they related?  The
connection we have made is simple: for problems in which the
low-temperature phase has multiple pure states (of non-negligible
thermodynamic weight) not related by spin inversion symmetry, the
broken ergodicity shows up as multiple extensive eigenvalues of the
spin-spin correlation matrix $C$. We have strong arguments in two
directions: first, that the presence of multiple extensive eigenvalues
necessarily implies nontrivial ergodicity breaking, i.e., multiple
pure-state pairs; and second, that the presence of multiple pure-state
pairs will give rise to multiple extensive eigenvalues.  We have found
striking support for these arguments from numerical (Monte-Carlo)
studies of the Sherrington-Kirkpatrick problem in three distinct
phases---the paramagnetic, ferromagnetic, and replica-symmetry-broken
(RSB) phases.  Specifically, we find clear and unambiguous signs of
the different kinds of ergodicity breaking in these three phases via
studies of the $N$ dependence of the disorder-averaged eigenvalues of
$C$---which essentially enable us to {\it count\/} the number of
extensive eigenvalues, and hence the number of pure-state pairs in the
configuration space.  We believe that the evidence for RSB displayed
in Figs.\ \ref{SK_ave}(a) and \ref{SK_typ} is unique in its
directness, clarity and lack of ambiguity.

We have also applied these ideas to the near-neighbour Ising spin glass in
four dimensions. Our data are consistent with the presence of only one
extensive eigenvalue, for $N\le 1296$. Furthermore, the typical value of
$\lambda_{2}/N$ decays with a clean power law; and the exponent agrees
roughly with the value expected from an argument based on the droplet
picture, plus independent estimates of the scaling exponent $\theta$. An   %J925
alternative analysis that assumes that a second extensive eigenvalue is    %J925
present with large finite size corrections cannot be completely excluded,  %J925
although the lack of any restriction on the fitting parameters makes any   %J925
conclusions drawn from such fits (in general of 
%%% HC1019
higher $\chi^2$) 
%%% lower $\chi$) 
questionable. %J925 %A103
Thus our results tend to support the `droplet' picture of the frozen
phase, with a single PSP, more strongly than they support the RSB picture. We
believe that studies of the kind reported here should be extended to larger
$N$ in order to test this tentative conclusion. Our present results
encourage us to believe that such studies can play an important role in
settling the question, from the theoretical side, of the nature of the broken
ergodicity in real spin glasses.

The authors acknowledge helpful discussions with J. Hu,
E. Sorensen, and G. Parisi. This work was supported by the National Science
Foundation under grants DMR-9820816 and DMR-9714055.

\appendix

\section{Properties of the matrices $C^0$, $V$ and $A$}
\label{app:M}
In this appendix we show that: (i) the symmetric matrices defined in   %%G1016
Sec.\ \ref{SEC:ssc}, $C^0$, $V$ (both of size $N \times N$) and $A$ (of size
$p \times p$), are positive semi-definite, (ii) the rank of $A$ is equal to   %%G1016
$\bar{p}$ (the number of pure states with linearly independent
magnetizations), (iii) the eigenvectors of $C^0$ constructed via   %%G1016
Eq.~(\ref{proposed}) from the linearly independent eigenvectors of $A$
with positive eigenvalue are linearly independent, and (iv) the   %%G1016
remaining $N-\bar{p}$ linearly independent eigenvectors of $C^0$ have
zero eigenvalue and are orthogonal to all pure state magnetizations.

We start by showing that $C^0$ is positive semi-definite: for an
arbitrary real N-dimensional vector $|v\rangle$,
\begin{eqnarray}
\sum_{ij} v_i C^0_{ij} v_j 
& = & \sum_{ij} v_i v_j  
\sum_{\alpha=1}^p w^\alpha m_i^\alpha m_j^\alpha \nonumber \\
& = & \sum_{\alpha=1}^p w^\alpha \Big( \sum_{i} v_i m_i^\alpha
\Big)^2 \ge 0.
\end{eqnarray}

Similarly, in the case of $V$ we have, 
\begin{eqnarray}
\lefteqn{ \sum_{ij} v_i V_{ij} v_j } \nonumber \\
& = & \sum_{ij} v_i v_j  
\sum_{\alpha=1}^p w^\alpha 
\langle (S_i - \langle S_i \rangle^{\alpha})
(S_j - \langle S_j \rangle^{\alpha}) \rangle^\alpha \nonumber \\
& = & \sum_{\alpha=1}^p w^\alpha \langle \Big( \sum_{i} v_i (S_i - \langle
S_i \rangle^{\alpha}) \Big)^2 \rangle^\alpha \ge 0.
\end{eqnarray}

We now concentrate in studying the $p \times p$ matrix $A$. By
convention, we enumerate the pure states so that the first $\bar p$
magnetization vectors $\{m^\alpha_j\}_{\alpha=1,\cdots,\bar{p}}$ are
linearly independent.  Now, for any real $p$-dimensional vectors 
$| x \rangle $ and $ | y \rangle $ we have,
\begin{eqnarray}
\langle x |A| y \rangle&=&
\sum_{\alpha,\beta=1}^p x^\alpha \sqrt{w^\alpha} q_{\alpha \beta} \sqrt{w^\beta}
y^\beta
\nonumber\\
&=& \frac{1}{N}\sum_{j=1}^N \left(\sum_\alpha^p x^\alpha \sqrt{w^\alpha}
m_j^\alpha\right)\left(\sum_\beta^p
y^\beta\sqrt{w^\beta}m_j^\beta\right). 
\label{xMy}
\end{eqnarray}
By choosing $ | x \rangle = | y \rangle = | v \rangle$ we immediately see that $A$ is positive
semi-definite. Thus statement (i) is proven.    %%G1016

We will now study the eigenvectors of the matrix $A$. Consider
Eq.~(\ref{xMy}) in the case that $ | x \rangle = | y \rangle = | v \rangle $
and $ | v \rangle $ is chosen such that
\begin{eqnarray}
v^{\alpha}=0 \ \ , \ \  \bar{p}<\alpha \le p  \nonumber \\
\sum_{\alpha=1}^p (v^\alpha)^2 >0 \ \ .
\label{whatisv}
\end{eqnarray}
Since by our 
%%% HC1019
assumptions $\{m^\alpha_j\}_{\alpha=1,\cdots,\bar{p}}$ are
linearly independent and $\{w^\alpha\}_{\alpha=1,\cdots,\bar{p}}$ are
nonzero, we have that 
%%%%%%%%%%%%%%%%%%%%%%%%%%%%%%%%%%%%%%%%%%%%%%%%%%%%%%%%%%%%%%%%%%%%%%%%
\begin{eqnarray}
\sum_{\alpha=1}^p v^\alpha\sqrt{w^\alpha}m_j^\alpha&=&
\sum_{\alpha=1}^{\bar{p}}v^\alpha\sqrt{w^\alpha}m_j^\alpha\ne 0\nonumber\\
\Rightarrow& & \langle v|A|v \rangle >0\,.
\label{EQ:A_pos}
\end{eqnarray}
By the Gram-Schmidt procedure, we can obtain $\bar{p}$ 
%%% HC1019
normalised
%%%
vectors $\{v_{r}\}$
that are 
%%% HC1019
orthogonal
%%% orthonormal 
with respect to $A$, all satisfying (\ref{whatisv}).
This means that 
\begin{equation}
\langle v_{r} | A | v_{s} \rangle = \delta_{r s} {a}_{r}
\label{EQ:Ars}
\end{equation}
where the $\{ {a}_{r} \}_{r=1,\cdots,\bar{p}}$ are positive
numbers. The $\{v_{r}\}$ will be shown below to be eigenvectors of
$A$ with eigenvalues ${a}_{r} > 0$, but first it is necessary to    %%G1016 
construct the eigenvectors with {\em zero} eigenvalue. To do that, let
us consider the magnetization vectors $m_j^\alpha$ with $\bar{p} <
\alpha \le p$, that can be written as linear combinations of the
first $\bar{p}$ ones (i.e. the ones that are linearly independent):
\begin{equation}
m_j^\alpha=\sum_{\beta=1}^{\bar{p}} z_\beta^\alpha m_j^\beta \ \ .
\label{EQ:lc_m}
\end{equation}
This allows to construct the $p - \bar{p}$ linearly independent
vectors $ | y_{s} \rangle $, $s=\bar{p}+1, \cdots, p$ of the form 
\begin{equation}
(y_{s})_\alpha 
= 
\left\{ 
\begin{array}{ll}
-\frac{z_\alpha^s}{\sqrt{w^\alpha}}, & 
\mbox{for $\alpha= 1,\cdots,\bar{p}$}, \\
\frac{\delta_{s \alpha}}{\sqrt{w^\alpha}} & 
\mbox{for $\alpha= \bar{p}+1,\cdots, p$}, 
\end{array}
\right.
\end{equation}
which satisfy
\begin{equation}
\sum_{\alpha = 1}^{p} (y_{s})_\alpha \sqrt{w^\alpha} m^\alpha_j = 0
\end{equation}
by Eq.~(\ref{EQ:lc_m}). Combining this with Eq.~(\ref{xMy}) it follows
that for an arbitrary 
$p$
dimensional real vector $| x \rangle$ 
\begin{equation}
\langle x| A | y_{s} \rangle = 0.
\label{EQ:xMy0}
\end{equation}
This means that all of the $ | y_{s} \rangle $ are eigenvectors of $A$ with
zero eigenvalue. By the Gram-Schmidt procedure, an orthonormal set $\{
| v_{\bar{p}+1} \rangle , \cdots, | v_{p} \rangle \}$ of linear combinations of them can be
constructed. By combining Eq.~(\ref{EQ:Ars}) and Eq.~(\ref{EQ:xMy0}) we get:
\begin{eqnarray}
A | v_{r} \rangle & = & {a}_{r} | v_{r} \rangle \qquad 
\mbox{for $r = 1, \cdots, p$}, 
\\
{a}_{r} & & 
\left\{ 
\begin{array}{ll}
> 0 & \mbox{for $ 1 \le r \le \bar{p} $}, \\
= 0 & \mbox{for $ \bar{p} < r \le p $},
\end{array}
\right.
\label{EQ:a_r}
\end{eqnarray}
and therefore statement (ii) is proven.    %%G1016

Consider now the vectors of the form proposed in Eq.~(\ref{proposed})
associated with the $\bar{p}$ eigenvectors of $A$ with positive
eigenvalue. Since the matrix $(v_{r})_{\alpha}$ ($r,\alpha = 1,
\cdots, \bar{p}$) is  invertible, and the
magnetization vectors $\{ m^{\alpha} \}_{\alpha = 1, \cdots, \bar{p}}$) are
linearly independent, it follows that the vectors $\{ \phi_{r} \}_{r = 1,
\cdots, \bar{p}}$ are linearly independent. Thus statement (iii) is   %%G1016
proven. 

The remaining $N-\bar{p}$ eigenvectors of $C^0$ can be constructed as
follows. Take an arbitrary vector $| u \rangle $ in the
$(N-\bar{p})$-dimensional subspace orthogonal to the magnetization
vectors $ \{ m^{\alpha} \}_{\alpha = 1, \cdots, \bar{p}} $ (and
consequently orthogonal to all of the $p$ magnetization vectors of the
pure states). Clearly, $| u \rangle $ is a null vector for $C^0$ since
\begin{equation}
\sum_{j} C^0_{ij} u_j = 
\sum_{\alpha=1}^p w^\alpha m_i^\alpha (\sum_j m_j^\alpha u_j) = 0, 
\end{equation}
and this proves statement (iv).   %%G1016

\section{Spectrum of $C^0 + V$}
\label{app:strong_clustering}
In this appendix some bounds are shown to be satisfied by the changes
in the eigenvalues of the spin-spin correlation matrix due to the
effect of $V$. 

For a given disorder realization, temperature $T$ and size $N$, let us consider
the eigenvalues and eigenvectors of the correlation matrices $C = C^0+V$ and
$C^0$: 
\begin{eqnarray} 
C |\chi_{r}\rangle & = & \lambda_{r} |\chi_{r}\rangle, \nonumber \\
C^0 |\phi_{r}\rangle & = & \kappa_{r} |\phi_{r}\rangle, 
\end{eqnarray}
where the eigenvalues are positive and labeled in descending order (the
subscript $1$ corresponding to the highest eigenvalue for each matrix). We
label by $\upsilon$ the largest eigenvalue of the matrix $V$.  We will use
variational arguments to show that:
\begin{itemize}

\item{} For any value of $\bar{p}$,
\begin{equation}
\kappa_{1} + \upsilon \ge \lambda_{r} \ge \kappa _{r} 
\mbox{ for $r = 1, \cdots, \bar{p}$}.
\label{EQ:lambda_lower}
\end{equation}

\item{} For $\bar{p} =1$, and if the ratio 
%%% HC1019 typo fixed
$\upsilon / \kappa_{1}$
%%% $\upsilon / \kappa_{\bar{p}}$ 
is small enough, then:
\begin{equation}
\lambda_{2} \le \upsilon (1 + \frac{\upsilon}{\kappa _{1}} +
O(({\frac{\upsilon}{\kappa _{1}}})^2) ).
\label{EQ:lambda_upper}
\end{equation}

\end{itemize}

Let's consider Eq.~(\ref{EQ:lambda_lower}) first. The upper bound is obvious
once $C$ is decomposed as a sum of $C^{0}$ and $V$, and use is made of the
fact that any mean value for each one of them has to be smaller or equal than
their respective maximum eigenvalues:
\begin{equation}
\lambda_{r} = \langle \chi_{r} | C | \chi_{r} \rangle 
            = \langle \chi_{r} | C^{0} | \chi_{r} \rangle
            +  \langle \chi_{r} | V | \chi_{r} \rangle 
          \le \kappa_{1} + \upsilon
\end{equation}

To prove the lower bound in the case $r=1$, we just consider the eigenvector
$ | \phi_{1} \rangle $ corresponding to the largest eigenvalue $\kappa_{1}$
of $C^{0}$, and use it as a variational trial vector:
\begin{equation}
\lambda_{1} \ge \langle \phi_{1} | C |  \phi_{1} \rangle = 
                \kappa_{1} + \langle \phi_{1} | V |  \phi_{1} \rangle 
            \ge \kappa_{1},
\end{equation}
where we have used the fact that $V$ is positive semi-definite.

To prove the lower bound for general $r = 1, \cdots, \bar{p}$, we use an
inductive reasoning. We assume that Eq.~(\ref{EQ:lambda_lower}) is valid for
all $r' = 1, \cdots, r-1$, and propose as a variational trial vector a linear
combination $| \tilde{\chi} \rangle$ of $ | \phi_{1} \rangle, \cdots, |
\phi_{r} \rangle $. Because it is generated by $r$ linearly independent
vectors, this linear combination can be chosen to be orthogonal to all of the
$r-1$ exact eigenvectors $| \chi_{1} \rangle, \cdots, | \chi_{r-1} \rangle$.
Then we have
\begin{equation}
\lambda_{r} \ge \langle \tilde{\chi}  | C | \tilde{\chi} \rangle = 
                \langle \tilde{\chi}  | C^{0} | \tilde{\chi} \rangle 
              + \langle \tilde{\chi}  | V | \tilde{\chi} \rangle 
            \ge \kappa_{r}, 
\end{equation}
where we have used the facts that i) the term corresponding to $C^{0}$ has
$\kappa_{r}$ as a lower bound, and ii) $V$ is positive semi-definite. 
This proves the inductive step, and therefore Eq.~(\ref{EQ:lambda_lower}).

Let's consider the inequality Eq.~(\ref{EQ:lambda_upper}). To prove it, we
decompose the eigenvector $|\chi_{1}\rangle$ into a part proportional to
$|\phi_{1}\rangle$ and a part orthogonal to it:
\begin{equation}
| \chi_{1} \rangle = \cos(\frac{\theta}{2}) | \phi_{1} \rangle +
                       \sin(\frac{\theta}{2}) | \eta \rangle,
\label{EQ:chi_{1}}
\end{equation}
%%% HC1019
where $ | \eta \rangle $ is a normalised vector orthogonal to $| \phi_{1}\rangle$.
%%%
We now write the matrix elements of $C$ in the subspace generated by $ | \chi_{1}
\rangle $ and $ | \eta \rangle$: 
\begin{equation}
\pmatrix{C} = 
\pmatrix{
\kappa_{1} + \langle \phi_{1} | V | \phi_{1} \rangle  & 
           \langle \phi_{1} | V | \eta \rangle \cr
           \langle \eta | V | \phi_{1} \rangle  & 
           \langle \eta | V | \eta \rangle \cr
        }\label{EQ:matrix_C_2x2}
\end{equation}
The coefficient $\theta$ that parametrizes Eq.~(\ref{EQ:chi_{1}}) can be
related to these matrix elements by
\begin{equation}
\tan \theta = \frac{ 2 \langle \phi_{1} | V | \eta \rangle } 
                   { \kappa_{1} + \langle \phi_{1} | V | \phi_{1} \rangle 
                              - \langle \eta | V | \eta \rangle },
\end{equation}
and therefore satisfies the bound 
\begin{equation}
| \tan \theta | \le \frac{ 2 \upsilon } { \kappa_{1} - \upsilon }.
\label{EQ:tan_bound} 
\end{equation}
We now consider the eigenvector $| \chi_{2} \rangle$, corresponding to
$\lambda_{2}$,  the second largest eigenvalue of $C$. Since $| \chi_{2} \rangle$
is normalized and orthogonal to $| \chi_{1} \rangle$, it has the form:
\begin{equation}
| \chi_{2} \rangle = 
  \cos(\frac{\phi}{2}) ( -\sin(\frac{\theta}{2}) | \phi_{1} \rangle +
                          \cos(\frac{\theta}{2}) | \eta \rangle )     
+ \sin(\frac{\phi}{2}) | \eta' \rangle, 
\end{equation}
where $ | \eta' \rangle $ is a normalized vector orthogonal both to $| \chi_{1}
\rangle$ and to $ | \eta \rangle $. From the expression for $\lambda_{2}$:
%%% HC1019 formula modified for clarity in first line 
%%% and to correct my typo in the second line
\begin{eqnarray}
  \lambda_{2} & = & \langle \chi_{2} | C | \chi_{2} \rangle  
            = \langle \chi_{2} | C^{0} | \chi_{2} \rangle
            +  \langle \chi_{2} | V | \chi_{2} \rangle \nonumber \\
              & = & \cos^{2}(\frac{\phi}{2}) \sin^{2}(\frac{\theta}{2}) 
                    \langle \phi_{1} | C^{0} | \phi_{1} \rangle + 
                    \langle \chi_{2} | V | \chi_{2} \rangle 
\end{eqnarray}
%%% HC1019 modified the last part to make it slightly more clear
we immediately obtain the bound:
\begin{equation}
  \lambda_{2} \le \sin^{2}(\frac{\theta}{2}) \kappa_{1} + \upsilon.
\label{EQ:l2_gb}
\end{equation}
From Eq.~(\ref{EQ:tan_bound}), it is clear that for $\upsilon \ll \kappa_{1}$
we have:
\begin{eqnarray}
\sin^{2}(\frac{\theta}{2}) & = &
\left[ \frac{\tan \theta}{2} ( 1 + O(\theta^{2}) ) \right]^{2} \nonumber \\
& = & \left[ \frac{\upsilon}{\kappa_{1}} 
( 1 + O (\frac{\upsilon}{\kappa_{1}}) ) \right]^{2} ,
\end{eqnarray}
which combined with Eq.~(\ref{EQ:l2_gb}) implies
Eq.~(\ref{EQ:lambda_upper}). 
%%%%%%%%%%%%%%%%%%%%%%%%%%%%%%%%%%%%%%%%%%%%%%%%%%%%%%%%%%%%%%%%%%%%%%%%
%%% we immediately obtain the bounds:
%%% \begin{eqnarray}
%%%   \lambda_{2} & \le & \sin^{2}(\frac{\theta}{2}) \kappa_{1} + \upsilon 
%%% \nonumber \\
%%%               & \le & (1 + (1 + \delta) \frac{\upsilon}{\kappa_{1}} ) \upsilon.
%%% \end{eqnarray}
%%% The second line, valid for any $\delta > 0 $ as long as the ratio
%%% $\upsilon / \kappa _{\bar{p}}$ is small enough, was obtained by combining
%%% Eq.~(\ref{EQ:tan_bound}) with the fact that $\lim_{\theta \to 0} {2
%%%   \sin(\frac{\theta}{2})}/ {\tan \theta} = 1 $, and it implies
%%% Eq.~(\ref{EQ:lambda_upper}).
%%%%%%%%%%%%%%%%%%%%%%%%%%%%%%%%%%%%%%%%%%%%%%%%%%%%%%%%%%%%%%%%%%%%%%%%

\begin{figure}
\epsfxsize=3.3in
\centerline{\epsffile{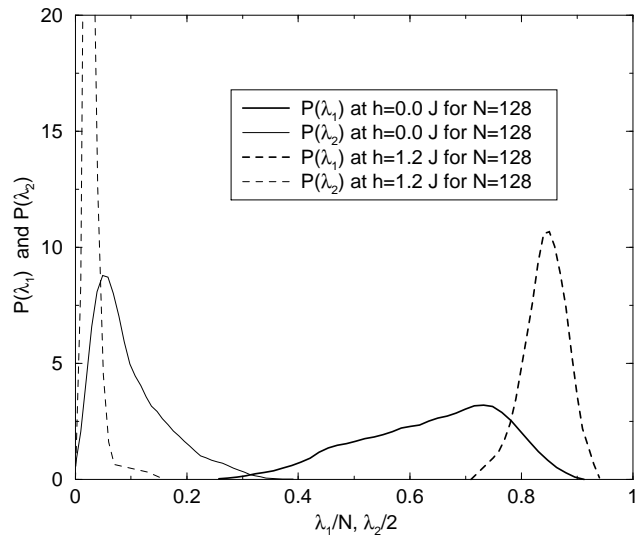}}
\caption{Distribution of the first (thick lines) and second (thin lines)
eigenvalues of $C_{ij}$  in the SK model at
$h/J=0$ (RSB phase; solid line) and $h/J=1.2$ (paramagnetic  %%G1016 
phase; dashed line).  %%G1016 
Here $T/J=0.4$ and $N=128$. These distributions, and the ones
shown in Fig.\ \ref{nicefit}, are a smoothed fit to binned data.}
\label{distrib}
\end{figure}

\begin{figure}
\epsfxsize=3.375in
\centerline{\epsffile{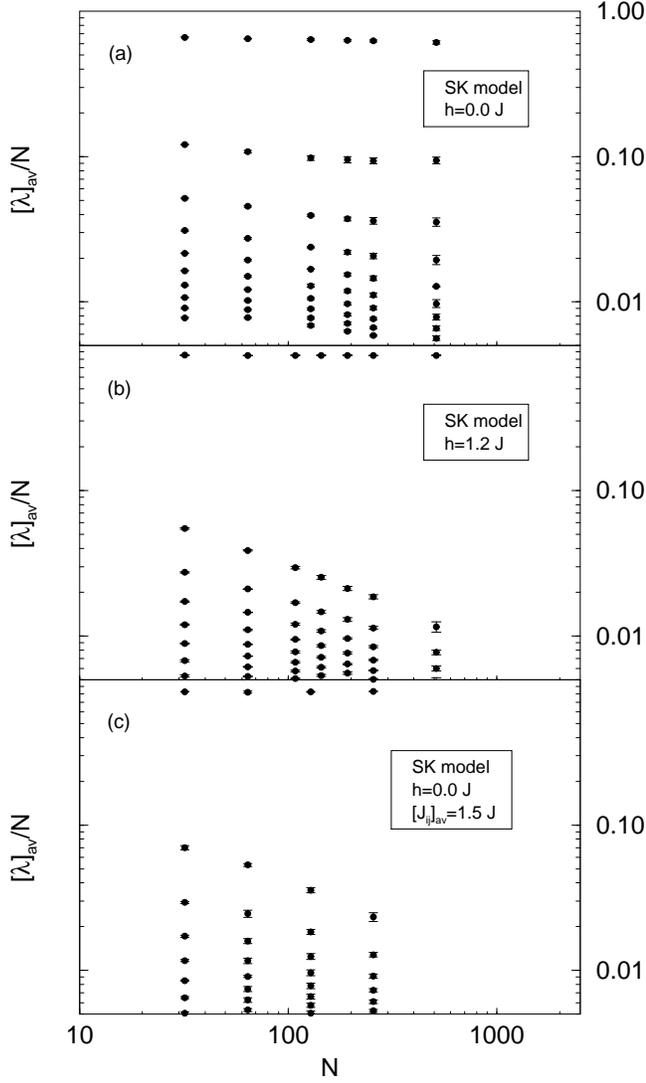}}
\caption{Scaling of the disorder average of the ten largest  %%G1016 
eigenvalues of $C_{ij}$  as a function of system size $N$ in the SK model 
(a) below the AT line ($h=0$, $T/J = 0.4$), (b) above the
AT line ($h/J=1.2$,  $T/J = 0.4$), and (c) in the ferromagnetic phase  %% G1016 
($h=0$, $T/J = 0.4$, $[J_{ij}]_{\rm av}=1.5 J$). The error bars in this
and subsequent figures come from the disorder averaging.}
\label{SK_ave}
\end{figure}

\begin{figure}
\epsfxsize=3.375in
\centerline{\epsffile{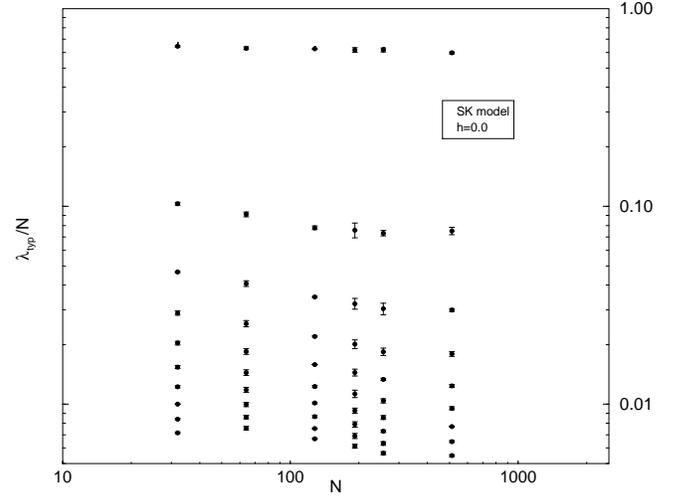}}
\caption{Scaling of the typical value ($[\lambda_i]_{\rm typ}   %%%G1016 
\equiv \exp[\ln \lambda_i]_{\rm av}$)
of the ten largest
eigenvalues of $C_{ij}$  as a function of system size $N$ in the SK model
below the AT line ($h=0$, $T/J = 0.4$).}
\label{SK_typ}
\end{figure}

\begin{figure}
\epsfxsize=3.375in
\centerline{\epsffile{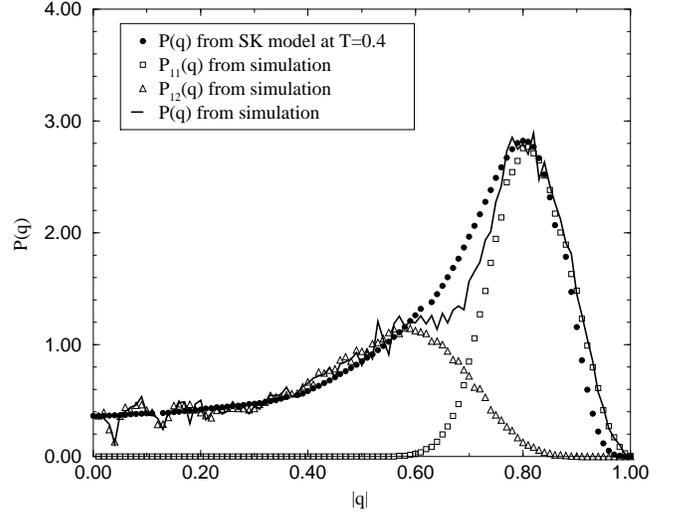}}
\caption{Overlap distribution function obtained from MC calculations in the
SK model at $T=0.4$ and $h=0$ (filled circles), self-overlap distribution function
$P_{11}(q)$ (open squares), mutual overlap distribution function
$P_{12}(q)$ (open triangles),  and the total $P(q)$ used
in the two-PSP model simulations (solid line).}
\label{overlap}
\end{figure}

\begin{figure}
\epsfxsize=3.375in
\centerline{\epsffile{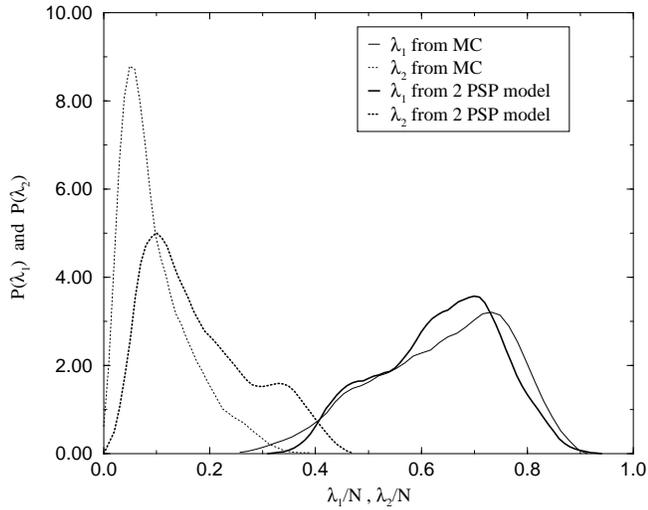}}
\caption{Distribution of the first (solid line) and second (dashed line)
eigenvalues of $C_{ij}$ obtained from the MC simulation of the SK model at
$N=64$ and $T/J=0.4$ (thin line), and the respective distributions
obtained from the two pure state model simulation (thick line) with
$w = 1/2$.}
\label{nicefit}
\end{figure}

\begin{figure}
\epsfxsize=3.375in
\centerline{\epsffile{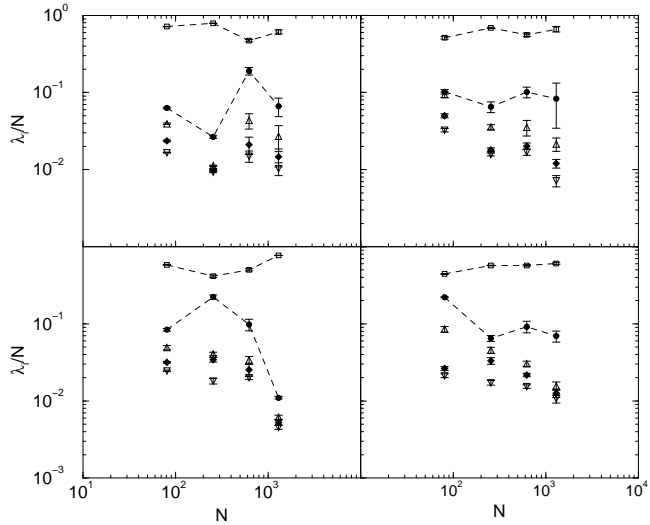}}
\caption{Four examples of the behavior of the $\{ \lambda_i/N \} $,
$1 \le i \le 5$, for a single disorder realization but increasing
$N$. Error bars are from Monte-Carlo fluctuations rather than
disorder averaging; dashed lines connecting $\lambda_{1}(N)$ points,
and $\lambda_{2}(N)$ points, are simply guides to the eye. 
We see that the behavior is very irregular, allowing no
conclusions about the number of extensive eigenvalues.}
\label{CSD}
\end{figure}

\begin{figure}
\epsfxsize=3.375in
\centerline{\epsffile{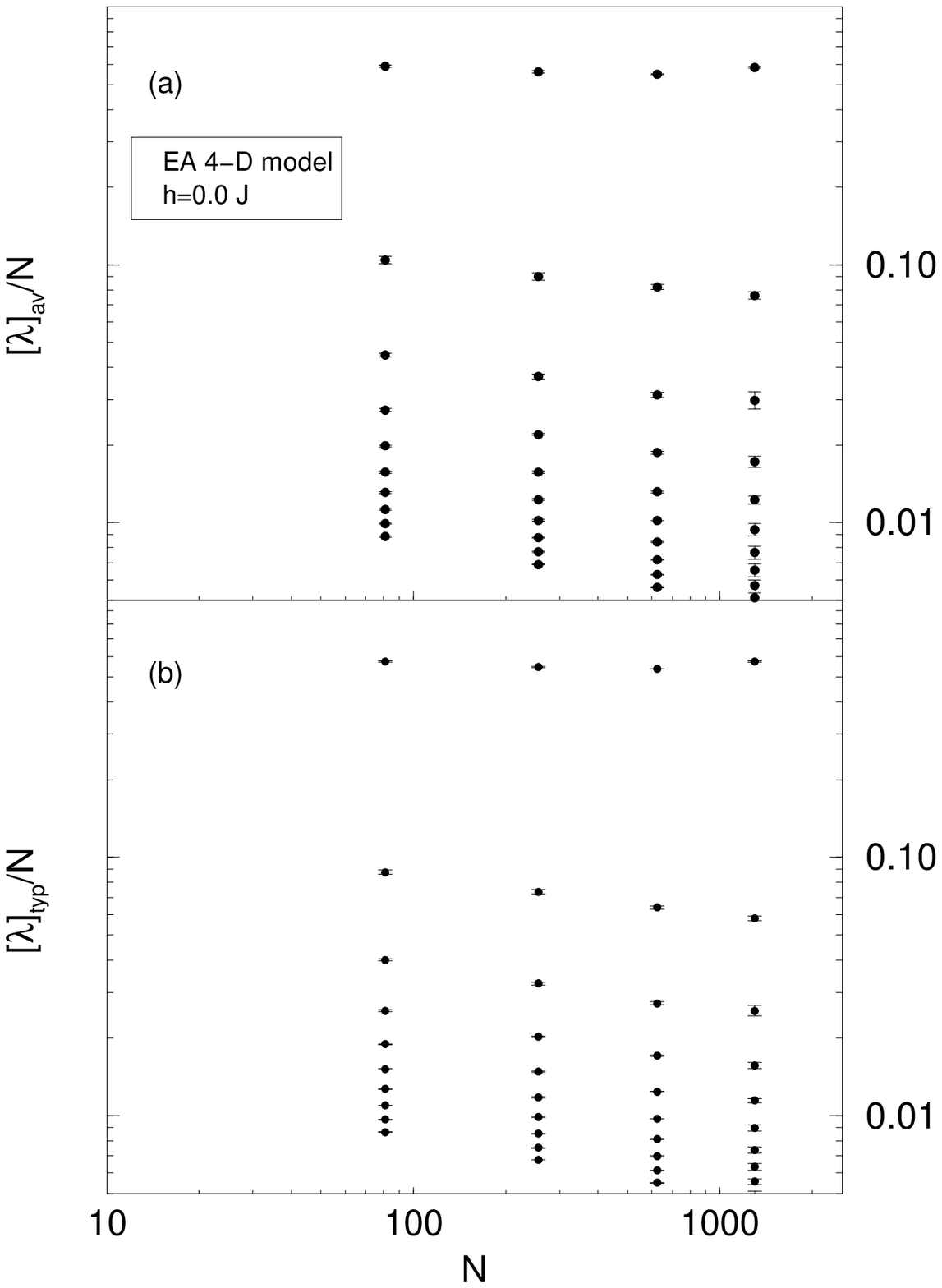}}
\caption{Scaling of the average (a) and the typical (b) value of the ten largest
eigenvalues of $C_{ij}$  as a function of system size $N$ for $h=0$ 
and $T/J=1.0$ for the EA model in four dimensions.}
\label{EA4D}
\end{figure}


\begin{references}
\bibitem{Par_comm} In this case $\chi^2=1.25$, which is still 
statistically acceptable, but a more important point is 
that given the small range of sizes several 
functional forms which would not correspond to a finite size correction 
also provide a similar fit.   %A103
\bibitem{FNOTE:one_state} To see this, first notice that $\langle S_i
S_j \rangle^- = (-1)^2 \langle S_i S_j \rangle^+ = \langle S_i S_j
\rangle^+ $ (the superscripts $+$ and $-$ respectively label
quantities associated with one of the pure states and with the other,
spin reversed, one). Suppose that $w^+$ and $w^- = 1-w^+$ are the
weights of the two pure states in the thermodynamic state of the
system; then
\begin{eqnarray*}
C & = & w^+ C^+ + (1-w^+) C^- \\
& = & (w^+ + 1-w^+ ) C^+ = C^+ = C^-,
\end{eqnarray*}
independently of the weights $w^+$ and $w^-$.
\bibitem{fse} Given our arguments that the extensive eigenvalues count the pure state pairs,
one need only assume that the leveling off of $[\lambda_2]_{\rm av}/N$ 
seen in Fig.\ \ref{SK_ave}(a)
will not reverse itself with larger $N$---a highly plausible assumption---in
order to conclude, without the need for further study at larger $N$,
that RSB is occurring.
\end{references}
\end{document}